# In-Situ Low-Angle Cross Sectioning: Bevel Slope Flattening due to Self-Alignment Effects

Uwe Scheithauer

*SIEMENS AG, CT MM 7, Otto-Hahn-Ring 6, 81739 München, Germany*

Phone: + 49 89 636 – 44143

E-mail: uwe.scheithauer@siemens.com

## Keywords

## Abstract

Low-angle cross sections are produced inside an Auger microprobe using the equipped depth profile ion sputter gun. Simply the sample is partly covered by a mask. Utilizing the edge of this mask the sample is sputtered with ions. Due to the shading of the mask a cross section is produced in the sample. The slope of this cross section is considerably shallower than given by the geometrical setup. This is attributed to self-alignment effects, which are due to missing sputter cascades in the transition area between sputtered and shaded sample regions and a chamfering of the mask edge.

These self-alignment effects are studied here using a 104.6 nm thick $SiO_2$ layer thermally grown on a Si substrate. In this study on one hand for a fixed ion impact angle of 15.8° as function of the sputter time several in-situ low-angle cross sections were produced. This way slope angles between an ultimate low slope angle of 0.014° and 0.085° were achieved. On the other hand for a fixed sputter time the ion impact angle was varied between 14.8° and 70.8°. For these samples cross section slope angles between 0.031° and 0.32° are observed. These results demonstrate the distinct slope flatting of in-situ cross sectioning.

## Introduction

A low-angle cross section can be produced inside an analytical Auger instrumentation using a mechanical mask and the depth profile ion sputter gun. Utilizing the edge of this mask the sample is sputtered inside the Auger microprobe with ions at nearly gracing incident angles of ~ 10° … 20°. This produces a cross section through the thin film system in the shadow of the mask with a very shallow slope, which is considerably flatter than given by the geometrical setup. Suitable mask materials are pieces of semiconductor wafers due to their good cleavage behavior or pieces of metal foils. Attributed to the clean ultra high vacuum environment the cross sections are contamination-free through the sputter removal.

Different methods of beveling are reported in the literature. Especially, mechanical and chemical approaches or methods, where a bevel is produced by varying the ion beam intensity laterally, were described. Mechanical plane bevels can be obtained by nearly surface parallel grinding using conventional metallography equipment **[1]**. More sophisticated approaches use either cylindrical **[2]** or spherical **[3-6]** erosion for this purpose. The latter one is often designated as ball cratering. For chemical beveling the upper sample layers are removed by an etchant. By varying the etch time laterally a bevel develops **[7]**. At the edge of an ion beam sputter crater, even if the ion beam is rastered, the ion intensity drops from maximum to zero producing a bevel. Its lateral dimensions are related to the ion beam intensity profile, the total ion dose and the sputter yield of the removed layers. To carry out crater-edge profiling mostly Auger line scans over the interesting bevels portion **[8-11]** or a conventional depth profile **[12]** applying alternating measuring and sputtering are recorded. For the conventional depth profile the measurement starts at a point where the interesting part of the thin film system is only covered by a thin top layer. The rough approach using the sputter crater edge suffers from the nonlinearity of the bevel sections. To get linear bevels the lateral ion beam dose can be controlled mechanically by a movable shutter **[10]** or by rastering the ion sputter beam electronically **[11]**. A little bit more special is elevated bevel sputtering **[13]**. On top of the sample a material ramp with a very shallow slope angle was deposited. Since this ramp has to be removed during sputtering, too, it reduces the effective ion dose etching the buried sample and leads to a bevel formation.

Up to now the benefits of using in-situ low-angle cross sections were demonstrated for two material analytic purposes **[12, 14, 15]**. On one hand, an in-situ low-angle cross section surface reveals the inner structure of samples up to some µm sample depth and gives an insight to the metallographic sample structure. The elemental composition can be determined by Auger measurements of interesting features. On the other hand, the use of in-situ low-angle cross sections for sputter depth profiling of polycrystalline thin film systems results in depth-resolution optimized depth profiles. Compared to sample rotation **[16, 17]**, the use of in-situ low-angle cross sections for depth-resolution optimized sputter depth profiling is more complex and time consuming. First, this effort is justified, because the depth-resolution of depth profiles using in-situ low-angle cross sections is better than that of sample rotation depth profiles if the layer thickness exceeds a certain material system dependent value. For polycrystalline Al layers, for instance, this value is in the range of 0.8 ... 1 µm **[14]**. Second, in-situ low-angle cross section depth profiling can be done on smaller features, as on metallization lines of microelectronic devices, for instance. Since no sample manipulator movements as for sample rotation depth profiling are necessary, this depth profile approach is not limited by the mechanical precision of the sample manipulator.

For thin film multilayer system containing polycrystalline metal layers, by in-situ low-angle cross sectioning using incoming ion angles in the range 10° … 20° slope angles between 0.3°... 3° were observed. This self-alignment effect of slope flattening was investigated here in detail by in-situ low-angle cross sectioning of a 104.6 nm thick $SiO_2$ layer thermally grown on a Si substrate. For the first series of cross sections the ion impact angle was fixed to 15.8° and the sputter time was varied. A second series was produced using ion impact angles between 14.8° and 70.8° and a fixed sputter time.

## Experimental

A Physical Electronics PHI 680 Auger microprobe was used for the Auger measurements presented here. The PHI 680 microprobe, an instrument with a hot field electron emitter, has a lateral resolution of ~ 30 nm at optimum. Under analytical working conditions using a higher primary current a lateral resolution of ~ 50 ... 70 nm is achievable. The Auger microprobe is equipped with differentially pumped $Ar^+$ ion sputter gun (model: 06-350). The instrument has a sample transfer system. The samples are mounted on sample holders, which are introduced to the main analysis chamber via a turbo pumped vacuum chamber.

All angles α are measured between the original sample surface and the incoming ion beam or slope of the cross section, respectively. So the angles relative to the surface normal are given by 90°- α.

## Bevel Flattening Mechanism of In-Situ Low-Angle Cross Sectioning

Fig.1 illustrates the basic concept of in-situ low-angle cross sectioning. The sample is partly covered by a mask. Utilizing the edge of this mask the sample is sputtered inside the Auger microprobe. Usually for analytical purpose incoming ions with an impact angle of approximately 10°... 20° are used **[12, 14, 15]**. For angle variation measurements presented here, the incoming ion angle was varied over a wider range. To ensure a homogeneous ion beam density, the ion beam is rastered electronically. That way in the shadow of the mask a cross section through the thin film system with a low angle is produced.

The slope of the in-situ low-angle cross section is considerably shallower than given by the geometrical setup. On one hand, the sputter process itself causes this. A surface atom will be sputtered by an incoming ion in a single collision process only with a low probability. More probably, the energy needed to remove a surface atom from the sample is transferred to it by a complex sputter cascade, which builds up in forward direction of the incoming ions. This has consequences for atoms positioned near to the shadow boundary between the sputtered and unsputtered area outside the shadow. Since in the shadowed area no sputter ions impinge the surface, for these atoms sputter cascades from the shadowed area are missed. Due to this, an atom in the shadow boundary proximity is sputtered less effectively than atoms far away from the shadow boundary. On the other hand, the edge of the mask is sputtered away during the in-situ low-angle cross section fabrication. The edge of the mask chamfers and thus the shadow boundary moves towards the mask. This reduces the total primary ion dose for the atoms near to the "moving" shadow boundary compared with atoms far away from it, because for these shadow boundary atoms sputtering starts with a certain time delay relative to the beginning of the in-situ low-angle cross section fabrication. Due to these two mechanisms the in-situ low-angle cross sections angle will only approximate to the incoming ions value after a very long sputter time, provided the mask withstands the sputtering long enough.

## Experimental Results for 104.6 nm $SiO_2$ on Si

The self-aligned slope flattening effects were investigated here by in-situ cross sectioning of a 104.6 nm thick $SiO_2$ layer thermally grown on a Si substrate. For the first experiment several $SiO_2$ samples were mounted on one sample holder using one large GaAs mask. This ensures the same experimental condition for all measurements and reduces experimental uncertainties. Uncertainties are due to the samples mount on the sample holder and the sample introduction process. For instance, due to particles on the sample a slit between sample and mask may establish. Or, while transferring the sample holder to the manipulator, due to the limited mechanical precision of the fitting, an uncertainty in angle is unavoidable. If all samples are mounted on one sample holder these limitations are less important, because the effect is similar for all samples.

For sputtering 2 keV $Ar^+$ ions were utilized. First a conventional depth profile was measured applying alternating measuring and sputtering with ions having the same impact angle as used for all further in-situ low-angle cross section preparations done on the $SiO_2$/Si samples. The sputter time $t_0$ needed to remove the $SiO_2$ layer was estimated. Thereafter six in-situ low-angle cross sections were produced using the same sputter conditions and sputter times t between $1.5t_0$ ... $91t_0$. An incoming ion beam angle of 15.8° was determined from the known mask thickness and the distance between mask and in-situ low-angle cross section. The lateral dimensions of the cross sections were measured using Auger linear scans (see fig. 2). On one side the $C_{KLL}$ signal decrease, which represents the surface contamination of the unsputtered shadow area, and on the other side the increase of the $Si_{KLL}$ substrate signal was taken for this. The inserts in fig. 1 and fig. 2 show secondary electron (= SE) images of the $4t_0$ in-situ low-angle-cross section. The position of the linear scan across the $4t_0$ in-situ low-angel cross section is marked in the SE image of fig. 2. The lateral dimension can be measured with an accuracy of approximately ± 2 % as tested by a magnification calibration reference.

Fig. 3 summarizes the results. The normalized lateral cross section length decreases and the slope of the cross section angle increases as function of the sputter time ratio $t/t_0$. These experimental results confirm the expectation discussed above. If longer sputter times are used for the in-situ low-angle cross section fabrications than the slope angle becomes steeper. For the $1.5t_0$ in-situ low-angle cross section the 104.6 nm thick $SiO_2$ layer is enlarged by a factor of ~ 4000 to 420 µm laterally. The ratio of the incoming ion beam angle to the slope angle of 0.014° is ~ 1110. These values confirm a tremendous self-alignment slope flattening effect for this thin film system.

In a second experiment for a fixed sputter time $4t_0$ five in-situ low-angle cross sections of the 104.6 nm thick $SiO_2$ were produced using different ion impact angles between 17.4° … 70.8°. The sputter time $t_0$ was estimated by an individual measurement of a conventional depth profile for each ion impact angle. If possible, again all samples were mounted on one sample holder using one larger mask. Only for the highest ion impact angle the sample was mounted on a special sample holder, which allows the extreme sample tilt. The results are shown in fig. 4. As function of the ion impact angle the normalized lateral cross section length and the cross section slope angle are plotted. With increasing ion impact angle the normalized lateral cross section length decreases and the cross section slope angle increases. But even for the highest ion impact angle of 70.8° the cross section slope angle is only ~ 0.32°, which gives a flattening ratio of ~ 220. Particularly with regard to an ion impact angle of 90°, which will produce a slope angle of 90°, this result is remarkable. Again, this series of experiments shows the tremendous self-alignment slope flattening effect of in-situ low-angle cross sectioning.

For the samples sputtered with ions having an impact angle of 33.8° the Ar implanted into the $SiO_2$ layer was measured via the peak-to-peak heights of the differentiated Auger signals. For this the Ar signal was normalized by the sum of the intensities of the Si, O and Ar signal. During the measurement of the conventional depth profile a normalized Ar proportion of 1.09 % was found. The Ar proportion of the $SiO_2$ layer of the in-situ low-angle cross section sample was estimated to 0.31 %. So the in-situ low-angle cross section sample has only ~ 1/3 of Ar implanted into the SiO2 layer. Supposable the $SiO_2$ layer of the in-situ low-angle cross section sample has less sputter induced defects than the $SiO_2$ layer of the conventional depth profile sample.

## Conclusions

For the material system 104.6 nm $SiO_2$ on Si the slope flattening self-alignment effect of in-situ low-angle cross sectioning was demonstrated by two measurement series. For a fixed ion impact angle of 15.8° the slope angle increases from 0.014° to 0.085° with the sputter time used for in-situ low-angle cross section fabrication. The 104.6 nm SiO2 layer was enlarged by a factor of 4000 to 420 μm laterally at maximum. By variation of the incoming ions impact angle even at steeper ion impact angles cross sections with shallow slopes are produced. For largest ion impact angle of 70.8° the cross section slope angle is only 0.32°. In summery, by these experiments cross sections with slope angles between 0.014° and 0.32° were obtained. The results of both measurement series demonstrate the distinct slope flatting of in-situ cross sectioning.

Both experiments confirm the presumptions of the slope flatting mechanisms discussed here. For atoms in close proximity to the shadow boundary outside the shadow sputter cascades from the shaded area are missed. And the shadow boundary itself moves towards the mask due to the masks edge chamfering. These mechanisms reduce the sputter efficiency and total ion dose, respectively, for the atoms near to the "moving" shadow boundary compared with atoms far away from it. The less effectively sputtered atoms retard sputtering of atoms, which are positioned behind them. This way an in-situ cross section slope develops, which is considerably shallower than given by the geometrical setup.

The usefulness of this bevel preparation method for the analysis of the cross section surface and for high depth-resolution depth profiling purpose has already been demonstrated. Further applications are conceivable. Since by in-situ low-angle cross sectioning bevels are produced, which have only a small inclination relative to the original sample surface, it may replace the mechanical sample preparation by grinding. Additionally, due to the ultra-high vacuum environment used, the in-situ low-angle cross sections are contamination free. In the future the in-situ low-angle cross sectioning may be used for preparation on vicinal surface, for example. In general, every technique, which uses such shallow bevels, can benefit from this bevel preparation technique.

### Acknowledgement

For fruitful discussions and invaluable suggestions I would like to express my thanks to my colleagues.

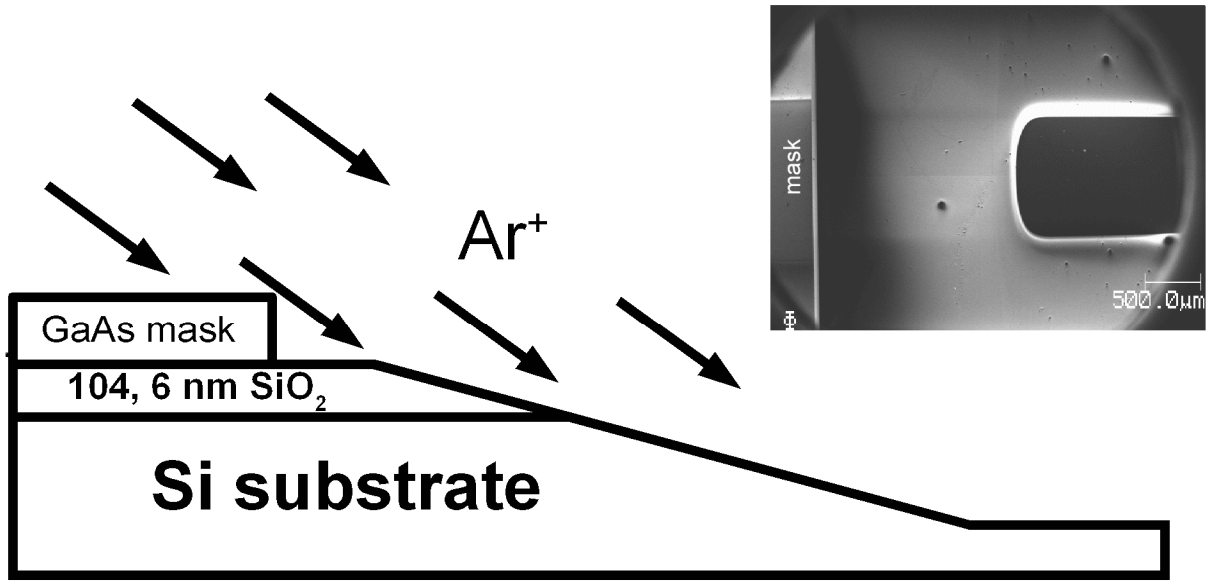

Fig. 1:   schematic diagram of an in-situ low-angle cross sectioning and a SE image of it
At the left side of the SE image the mask and at the right side the sputter crater in the SiO$_2$/Si sample is visible. The mechanical drift tube of the primary electron beam restricts the field of view.

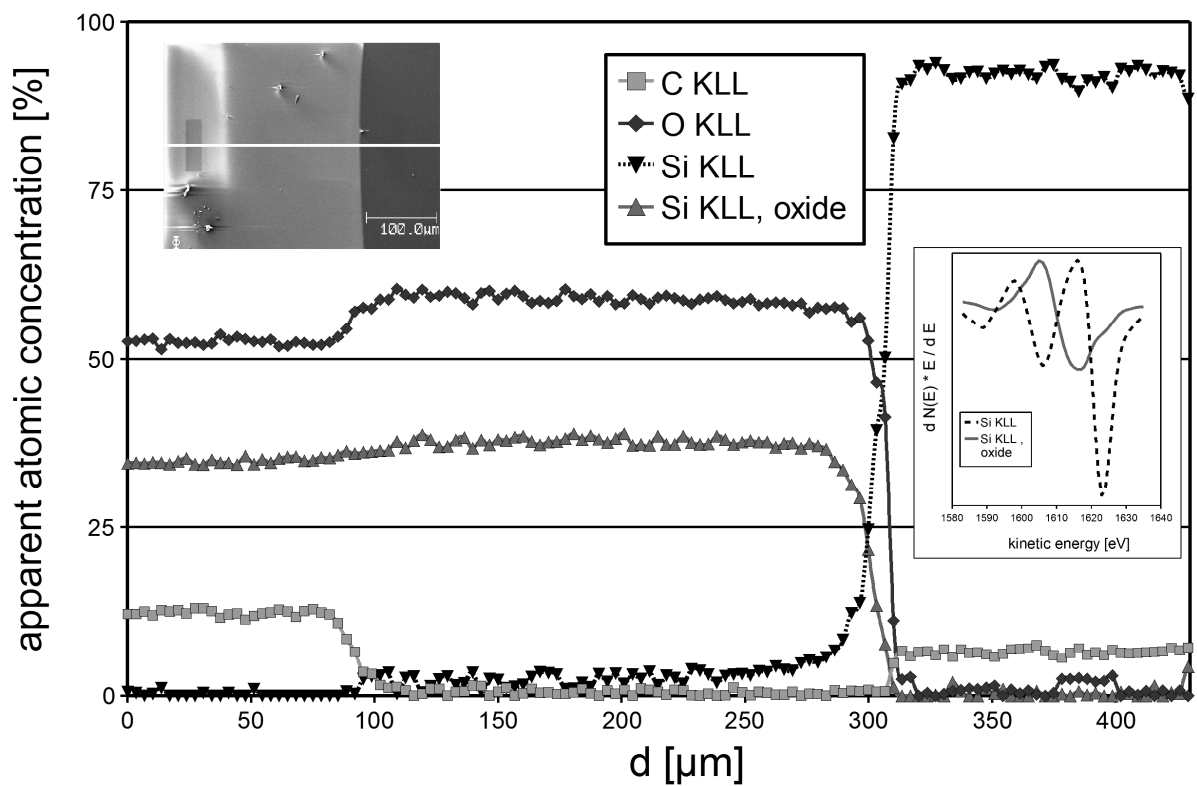

Fig. 2:   linear scan along an in-situ low-angle cross section
sputter time for in-situ low-angle cross section fabrication: $4t_o$
The position of the linear scan is marked in the inserted SE image. Peak fitting using the two different Si$_{KLL}$ signals deconvolutes Si and Si oxide.

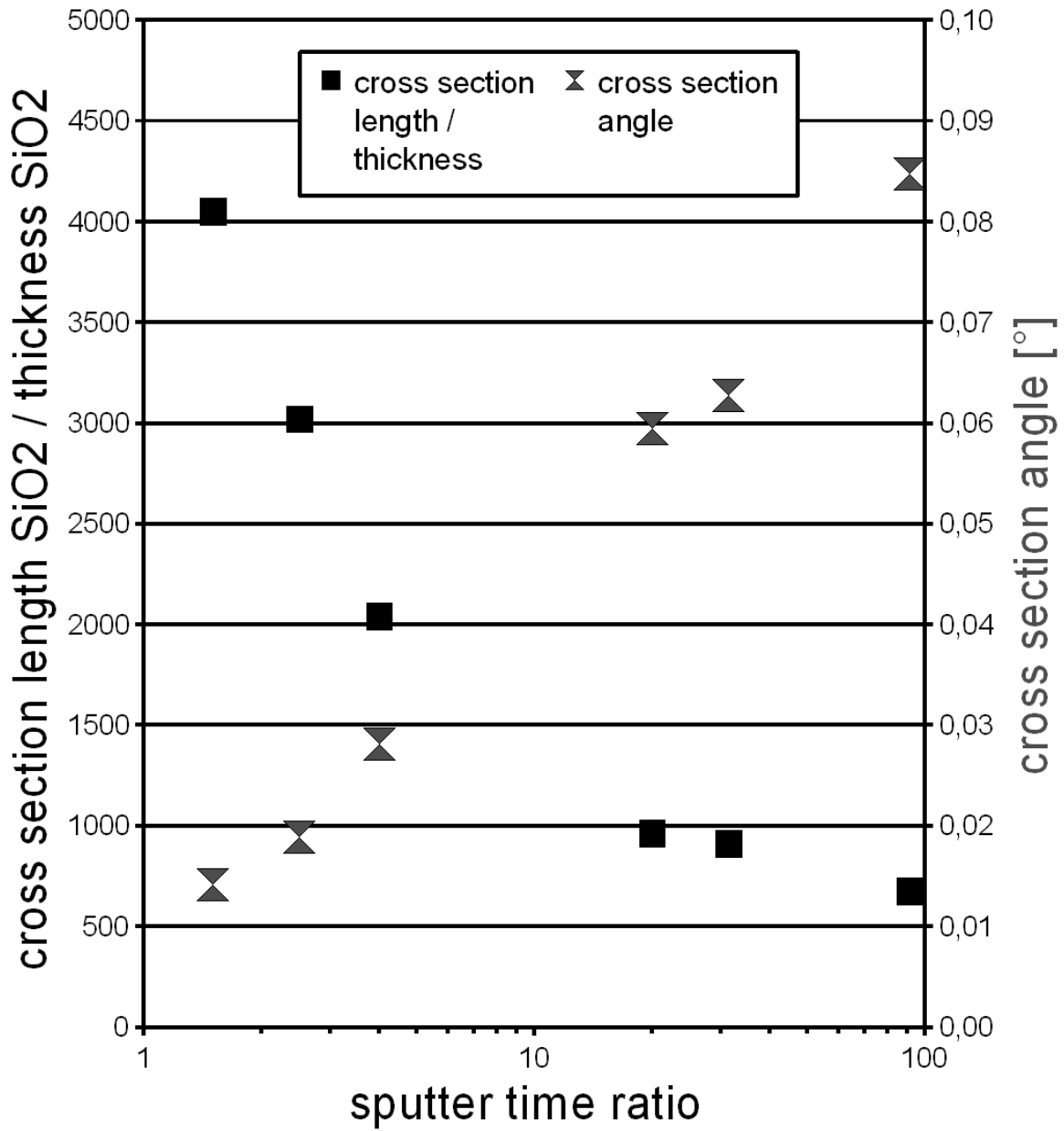

Fig. 3: cross-section length to thickness ratio and cross section angle, respectively, as function of sputter time ratio

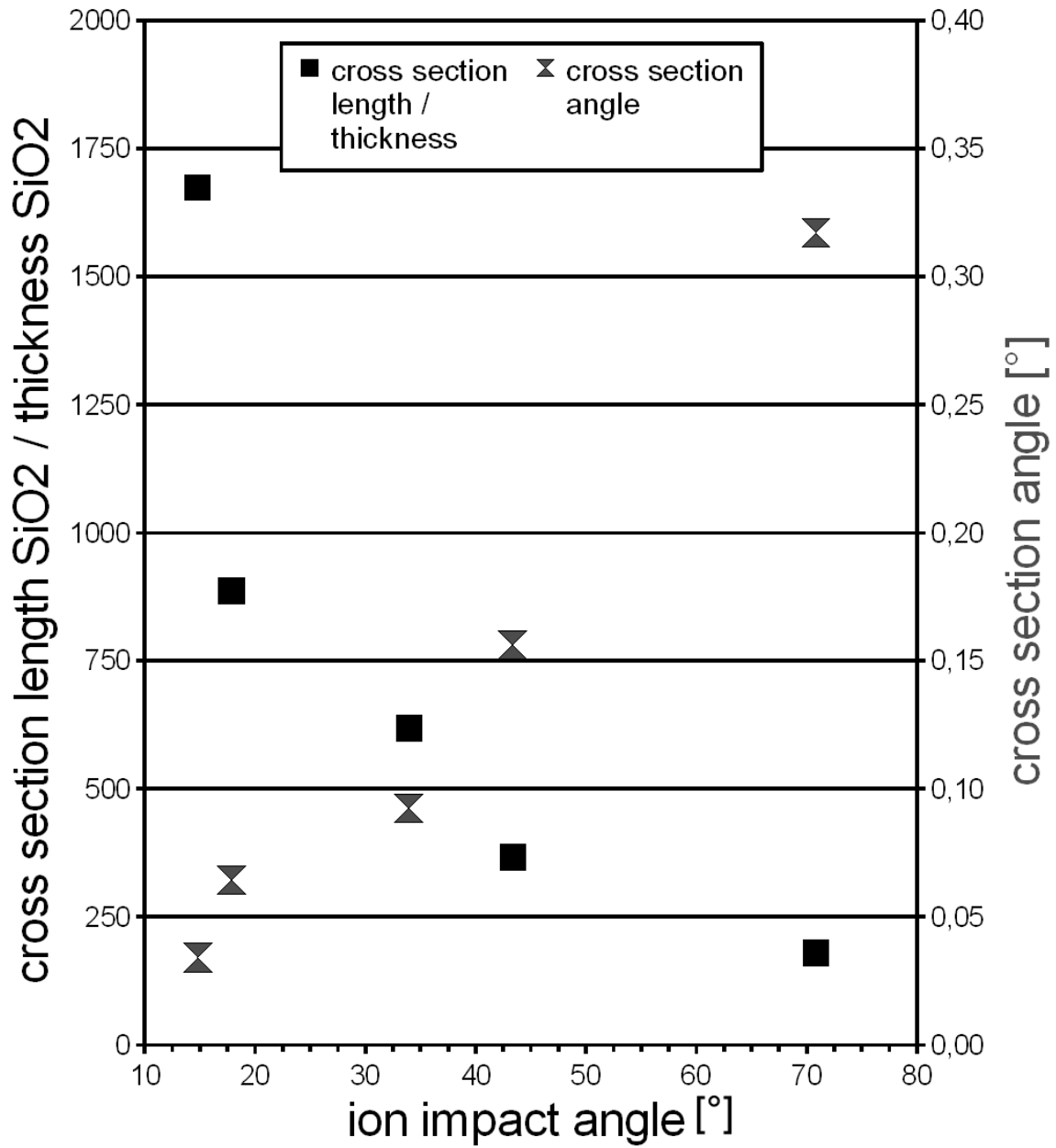

Fig. 4: cross-section length to thickness ratio and cross section angle, respectively, as function of ion impact angle